**Title**

Wearable intelligent throat enables natural speech in stroke patients with dysarthria

**Authors**

Chenyu Tang[†1], Shuo Gao[†*2], Cong Li[†2], Wentian Yi[1], Yuxuan Jin[3], Xiaoxue Zhai[4], Sixuan Lei[5], Hongbei Meng[2], Zibo Zhang[1], Muzi Xu[1], Shengbo Wang[2], Xuhang Chen[6], Chenxi Wang[2], Hongyun Yang[2], Ningli Wang[7], Wenyu Wang[8], Jin Cao[9], Xiaodong Feng[10], Peter Smielewski[6], Yu Pan[4], Wenhui Song[11], Martin Birchall[12], and Luigi G. Occhipinti[*1]

**Affiliations**

[1]Department of Engineering, University of Cambridge, Cambridge, UK

[2]School of Instrumentation and Optoelectronic Engineering, Beihang University, Beijing, China

[3]Cavendish Laboratory, University of Cambridge, Cambridge, UK

[4]Department of Rehabilitation Medicine, Beijing Tsinghua Changgung Hospital, Tsinghua University, Beijing, China

[5]Shenzhen International Graduate School, Tsinghua University, Shenzhen, China

[6]Department of Clinical Neurosciences, University of Cambridge, Cambridge, UK

[7]Beijing Tongren Hospital, Capital Medical University, Beijing, China

[8]Thrust of Smart Manufacturing, Hong Kong University of Science and Technology (Guangzhou)

[9]School of Life Sciences, Beijing University of Chinese Medicine, Beijing, China

[10]Department of Rehabilitation Center, The First Affiliated Hospital of Henan University of Chinese Medicine, Zhengzhou, China

[11]Department of Surgical Biotechnology, University College London, London, United Kingdom

[12]Royal National Ear Nose and Throat and Eastman Dental Hospitals, University College London Hospital, London, UK

[†]These authors contributed equally: Chenyu Tang, Shuo Gao, Cong Li
[*]Correspondence to: Shuo Gao (shuo_gao@buaa.edu.cn) and Luigi G. Occhipinti (lgo23@cam.ac.uk)

**Abstract**

Wearable silent speech systems hold significant potential for restoring communication in patients with speech impairments. However, seamless, coherent speech remains elusive, and clinical efficacy is still unproven. Here, we present an AI-driven intelligent throat (IT) system that integrates throat muscle vibrations and carotid pulse signal sensors with large language model (LLM) processing to enable fluent, emotionally expressive communication. The system utilizes ultrasensitive textile strain sensors to capture high-quality signals from the neck area and supports token-level processing for real-time, continuous speech decoding, enabling seamless, delay-free communication. In tests with five stroke patients with dysarthria, IT's LLM agents intelligently corrected token errors and enriched sentence-level emotional and logical coherence, achieving low error rates (4.2% word error rate, 2.9% sentence error rate) and a 55% increase in user satisfaction. This work establishes a portable, intuitive communication platform for patients with dysarthria with the potential to be applied broadly across different neurological conditions and in multi-language support systems.

## I. Main

Neurological diseases such as stroke, amyotrophic lateral sclerosis (ALS), and Parkinson's disease frequently result in dysarthria—a severe motor-speech disorder that compromises neuromuscular control over the vocal tract. This impairment drastically restricts effective communication, lowers quality of life, substantially impedes the rehabilitation process, and can even lead to severe psychological issues [1, 2, 3, 4]. Augmentative and alternative communication (AAC) technologies have been developed to address these challenges, including letter-by-letter spelling systems utilizing head or eye tracking [5, 6, 7, 8] and neuroprosthetics powered by brain-computer interface (BCI) devices [9, 10, 11, 12]. While head or eye tracking systems are relatively straightforward to implement, they suffer from slow communication speeds. Neuroprosthetics, while transformative for severe paralysis cases, often rely on invasive, complex recordings and processing of neural signals. For individuals retaining partial control over laryngeal or facial muscles, a strong need remains for solutions that are more intuitive and portable (Note S1).

A promising solution lies in wearable silent speech devices that capture non-acoustic signals, such as subtle skin vibrations [13, 14, 15, 16, 17] or electrophysiological signals from the speech motor cortex [18, 19, 20, 21]. These technologies offer non-invasiveness, comfort, and portability, with potential for seamless daily integration. Yet, despite their promise, current systems remain in their infancy, achieving reliable, discrete word decoding in healthy users but showing limited success in patient trials [13, 14, 15]. More critically, these systems fall short of delivering truly natural communication—requiring both delay-free expression and consistent contextual coherence, capabilities essential for fully effective and meaningful interactions.

To advance wearable silent speech systems for real-world dysarthria patient use, we developed an AI-driven intelligent throat (IT) system that captures extrinsic laryngeal muscle vibrations and carotid pulse signals, integrating silent speech and emotional states analysis in real-time. The system generates personalized, contextually appropriate sentences that accurately reflect patients' intended meaning (Figure 1). It employs ultrasensitive textile strain sensors, fabricated using advanced printing techniques, to ensure comfortable, durable, and high-quality signal acquisition [14, 22]. By analyzing speech signals at the token level (~100ms), our approach outperforms traditional time-window methods, enabling continuous, fluent word and sentence expression in real time. Knowledge distillation further reduces computational latency by 76%, significantly enhancing communication fluidity. Large language models (LLMs) serve as intelligent agents, automatically correcting token classification errors and generating personalized, context-aware speech by integrating emotional states and environmental cues. Pre-trained on a dataset from 10 healthy individuals, the system achieved a word error rate (WER) of 4.2% and a sentence error rate (SER) of 2.9% when fine-tuned on data from five dysarthric stroke patients. Additionally, the integration of emotional states and contextual cues further personalizes and enriches the decoded sentences, resulting in a 55% increase in user satisfaction and enabling dysarthria patients to communicate with fluency and naturalness comparable to that of healthy individuals. Table S1 provides a comprehensive comparison between the IT system and state-of-the-art wearable silent speech systems.

## II. Results

**The intelligent throat system**

The IT system consists primarily of hardware (a smart choker embedding textile strain sensors and a wireless readout printed circuit board (PCB)) and software components (machine learning models and LLM agents). Silent speech signals generated in real time by the user's silent expressions (silently mouthed words in the absence of vocalized sound) are decoded by a token decoding network and synthesized into an initial sentence by the token synthesis agent (TSA). Simultaneously, pulse signals are collected from the smart choker device and processed by an emotion decoding network to determine the user's real-time emotional status. The sentence expansion agent (SEA) intelligently expands the TSA-generated sentence, incorporating personalized emotion labels and objective contextual background data to produce a refined, emotionally expressive, and logically coherent sentence that captures the user's intended meaning (Fig. 1, Movie S2). Each component of the IT system is elaborated upon in the following sections.

Fig. 2A shows the structure of the strain sensing choker screen-printed on an elastic knitted textile. The choker features two channels located at the front and side of the neck, designed to monitor the strain applied to the skin by the muscles near the throat and the carotid artery (Fig. S1). The graphene layer printed on the textile forms ordered cracks along the stress concentration areas of the textile lattice to detect subtle skin vibrations [14]. Silver electrodes are connected to the integrated PCB on the choker. A rigid strain isolation layer with high Young's modulus is printed around each channel to reduce crosstalk between the two channels and the variable strains caused by wearing. Due to the difference in Young's modulus between the elastic textile substrate and the strain isolation layer, less than 1% of external strain is transmitted to the interior when wearing the choker, while the internal sensing areas remain soft and elastic (Fig. S2) [22]. For uniaxial stretching (x-axis) from 1-10 Hz, the printed textile-based graphene strain sensor shows good linear behavior, producing a response over 10% to subtle strains of 0.1% and maintaining a gauge factor (GF) over 100 during high-frequency stretching (Fig. 2B), while y- and z-axis deformations contribute negligible signal variations due to the anisotropic crack propagation mechanism. Based on our previous findings and related studies, the 0.1% strain threshold has been validated as sufficient for capturing silent speech-induced muscle vibrations [14, 15, 17]. Furthermore, our previous studies have confirmed the reliability of the printed textile-based strain sensors with high robustness, durability and washability, as well as high levels of comfort, biocompatibility and breathability [14, 22].

To operate the system and enable wireless communication between the IT choker and server, the PCB was designed for bi-channel measurements (i.e., silent speech and carotid pulse signals), enabling simultaneous acquisition of speech and emotional cues. The PCB integrates a low-power Bluetooth module (Fig. 2C) for continuous data transmission while optimizing energy efficiency for extended use. Key components of the PCB include an analog-to-digital converter (ADC) for high-fidelity signal digitization and a microcontroller unit (MCU) that manages data processing and transmission (Fig. 2D, Fig. S4, and Fig. S5). Power supply, operational amplifiers, and the reference voltage chip are configured to ensure stable signal amplification, catering to the sensitivity requirements of both strain and pulse sensors. For the energy management system, a comprehensive power budget analysis reveals that the designed PCB operates with a total power consumption of 76.5 mW (Fig. 2E). The main power-consuming components are the Bluetooth module (29.7 mW) and amplification circuits (31.9 mW). To extend operational time and support portable use, a 1800 mWh battery was incorporated, providing sufficient capacity for continuous operation thoughout an entire day without recharging.

**Token-level speech decoding**

Current wearable silent speech systems operate by recognizing discrete words or predefined sentences and lack the ability for continuous, real-time expression analysis typical of the human brain [41]. This limitation arises because these systems rely on fixed time windows (typically 1–3 seconds) for word decoding, requiring users to complete each word within a set interval and pause until the next window to continue [13-21]. Such constraints lead to fragmented expression and unnatural user experience. To address this, we developed a high-resolution tokenization method for signal segmentation (Fig. 2F), dividing speech signals into fine-grained ~100ms segments for continuous word label recognition. This granular segmentation ensures that each token accurately corresponds to a specific part of a single word and is labeled accordingly. This setup enables users to speak fluidly without worrying about timing constraints, as the system continuously classifies and aggregates tokens into coherent words and sentences. Our optimization determined that a token length of 144 ms offers the ideal balance: it minimizes boundary confusion from longer tokens while avoiding the increased computational demands associated with shorter tokens. This fine-grained segmentation not only eliminates the unnatural pauses imposed by prior fixed-time-window methods but also ensures that each token retains essential local signal features. Compared to traditional silent speech decoding approaches, which rely on whole-word classification, this token-based approach enables a real-time, continuous speech experience that more closely mimics natural spoken language.

While high-resolution tokenization improves fluidity, shorter tokens inherently contain limited context, making them less effective for accurate word decoding. Temporal machine learning models, like recurrent neural networks (RNN) or transformers, could capture contextual dependencies, but their complexity and computational cost render them suboptimal for wearable silent speech systems [23, 24, 25], which prioritize real-time operation. To balance context awareness and computational efficiency, we implemented an explicit context augmentation strategy (Fig. 3A), where each sample consists of N tokens: N-1 preceding tokens provide context, and the current token determines the sample's label. For initial tokens, any missing preceding tokens are padded with blank tokens to ensure completeness. We found N=15 tokens to be optimal (Fig. 3C), with accuracy initially increasing as tokens accumulate, then declining due to insufficient context at lower counts and gradient decay or information loss at higher counts [26]. This strategy enables the use of efficient one-dimensional convolutional neural networks (1D-CNNs) instead of computationally intensive temporal models for token decoding [27, 28]. Attention maps reveal that signals from preceding regions indeed contribute to token decoding, validating the effectiveness of the explicit context augmentation strategy (Fig. S10).

To further enhance model efficiency and accuracy on patients' data, we designed the training pipeline shown in Fig. 3B. The model was pre-trained on a larger dataset from healthy individuals and then fine-tuned on the limited patients' data, leveraging shared signal features to enhance patient-specific decoding. After only 25 repetitions per word in few-shot learning, the model achieved a token classification accuracy of 92.2% (Fig. 3D). In contrast, a model trained from scratch using solely patients' data could only reach an accuracy of 79.8%. Additionally, we employed response-based knowledge distillation [29] to transfer knowledge from a larger 1D ResNet-101 model to a smaller 1D ResNet-18, reducing computational load by 75.6% while maintaining high accuracy, with only a 0.9% drop from the teacher model, achieving 91.3% (Fig. 3E). Fig. 3F and Fig. 3G display the confusion matrix and UMAP feature visualization for token decoding [30]. Over 90% of the classification errors involved confusion between class 0 (blank tokens) and neighbouring word tokens. As shown in later analyses of the LLM agent's performance, such boundary errors can be effectively corrected during token-to-word synthesis by the token synthesis agent (TSA). This knowledge distillation and transfer learning framework ensures that computational efficiency is maximized without sacrificing accuracy. Unlike prior approaches that train models from scratch on small patient datasets, our pipeline generalizes well across individuals, addressing a key challenge in real-world silent speech decoding for dysarthric patients.

**Decoding of emotional states**

To enrich sentence coherence by providing emotional context, we decode emotional states from carotid pulse signals. Emotional state recognition can typically be achieved through a variety of methods, including analysis of facial images from cameras, audio speech signals, and various physiological indicators such as heart rate and blood pressure [31, 32, 33]. In line with our objective of creating a highly integrated wearable system, we chose carotid pulse signals as a biomarker for emotional decoding. Using 5-second windows, we segmented patients' pulse signals into samples to construct a dataset, focusing on three common emotion categories for stroke patients: neutral, relieved, and frustrated (data collection protocol detailed in Methods). Fig. 4A shows the discrete Fourier transform (DFT) distributions for each emotion, highlighting distinct frequency characteristics among these emotional states. Accordingly, we incorporated DFT frequency extraction into the decoding pipeline shown in Fig. 4B, where removal of the DC component, Z-score normalization, and DFT are sequentially applied before feeding the values into a classifier for categorization. The DFT-based approach was selected for its ability to represent key characteristics of carotid pulse signals, including power distribution, frequency-domain features, and waveform morphology, within a single transformation. This method enables our end-to-end neural network to automatically extract the most relevant features for emotion classification, eliminating the need for manual feature engineering. Fig. 4C illustrates the performance of different classifiers with and without DFT frequency extraction. The results show a significant improvement in decoding accuracy with DFT. The optimal model was the 1D-CNN with DFT, achieving an accuracy of 83.2%, with its confusion matrix displayed in Fig. 4D. The SHAP values reveal that the emotion decoding model primarily focuses on low-frequency signals in the 0-2 Hz range, which is consistent with the pulse signal range demonstrated by the DFT (Fig. S11).

In addition to the silent speech and carotid pulse signals analyzed in this study, various physiological activities generate distinct vibrational signals in the neck area, which can introduce artefacts hindering analysis [34, 35]. Fig. 4E shows the frequency and magnitude distributions of several prominent signals in this region. Our observations revealed that silent speech exhibits a relatively strong magnitude, and in applications with the IT, vibration can propagate transversely from the throat center to the carotid artery, introducing crosstalk in the pulse signal. Due to the considerable frequency overlap between silent speech and pulse signals, digital filters are non-ideal for effective artefacts suppression [36]. While adding reference channels could theoretically help, it does not align with the goal of a highly integrated IT [37]. To address this issue, we employed a stress isolation treatment using a polyurethane acrylate (PUA) layer, as shown in Fig. 2a, to prevent strain crosstalk propagation along the IT. The theoretical basis of this isolation strategy has been thoroughly discussed in our previous study [22]. Fig. 4F compares pulse signals with and without strain isolation treatment when silent speech occurs concurrently (the vowel "a" introduced at 2.5s), demonstrating significant crosstalk resilience in the treated IT.

**LLM agents for sentence synthesis and intelligent expansion**

To naturally and coherently synthesize sentences that accurately reflect the patient's intended expression from the decoded token and emotion labels, we introduced two LLM agents based on the GPT-4o-mini API (Fig. 5A): the token synthesis agent (TSA) and the sentence expansion agent (SEA). The TSA merges token labels directly into words silently expressed by the patient and combines them into sentences (left). The SEA, on the other hand, leverages emotion labels and objective information, such as time and weather, to expand these basic sentences into logically coherent, personalized expressions that better capture the patient's true intent. Through a simple interaction (in this study, two consecutive nods), the IT system enables seamless switching between the direct output and the enriched, expanded sentence.

To optimize the performance of the TSA, we refined the prompt design [38]. First, we optimized the prompt length (Fig. 5B), observing a trend where both WER and SER improved with increasing prompt length up to 400 words before eventually deteriorating for higher lengths. We attribute this trend to the fact that longer prompts provide clearer synthesis instructions, but overly lengthy prompts dilute the model's focus ability. Additionally, we compared performance with and without example cases, where the agent was provided with five examples of token label sequences and their corrected word outputs. Including examples significantly improved synthesis accuracy (Fig. 5C). Finally, we evaluated the effect of providing empirical constraints, which specify typical token counts for words of various lengths. Performance improved considerably when constraints were included (Fig. 5D). Under optimal prompt conditions, TSA achieved its best performance with a WER of 4.2% and an SER of 2.9%.

We also assessed and refined the performance of the SEA. Patient satisfaction with the expanded sentences was evaluated through a questionnaire (see Table S4 for criteria details). Following Chain-of-Thought (CoT) optimization [39] and the inclusion of patient-provided expansion examples, the expanded sentences scored significantly higher across multiple criteria (Fig. 5F). Contribution analysis revealed that emotion labels made a substantial impact on emotion accuracy, while objective information notably improved fluency, jointly contributing to the overall satisfaction with the expanded sentences compared to the basic word-only output (Fig. 5E). Under optimal prompt conditions, the SEA-generated expanded sentences resulted in a 55% increase in overall patient satisfaction compared to the TSA's direct output, raising satisfaction from "somewhat satisfied" to "fully satisfied" levels (Fig. S12 and Fig. S13).

As shown in Fig. 5F, the core meaning metric remains stable across all sentence expansion conditions. This stability stems from the high accuracy of the token decoding model and TSA, which ensure precise word recognition and correct token synthesis. Since core meaning reflects whether the fundamental subject-verb-object (SVO) structure aligns with the user's intended message, this metric remains largely unchanged after expansion. However, as illustrated in Fig. 5E, additional contextual information - including objective data (e.g., time, weather) and emotion labels - enriches fluency and personalization, significantly improving overall user satisfaction. In both operating modes, sentences generated by the TSA and SEA agents are sent to an open-source text-to-speech model [40], which synthesizes audio that matches the patient's natural voice for playback. In real-world applications, the delay between the completion of the user's silent expression and the sentence playback is approximately 1 second (Note S2). This low latency effectively supports seamless and natural communication in practical settings. To assess the long-term adaptability of the IT system, we conducted a follow-up test six months after initial training, observing an increase in WER due to changes in neuromuscular control, which was rapidly restored to initial performance levels after a brief few-shot fine-tuning (five repetition per words) session (Table S5).

## III. Discussion

In this work, we introduce the IT, an advanced wearable system designed to empower dysarthric stroke patients to communicate with the fluidity, intuitiveness, and expressiveness of natural speech. Comprehensive analysis and user feedback affirm the IT's high performance in fluency, accuracy, emotional expressiveness, and personalization. This success is rooted in its innovative design: ultrasensitive textile strain sensors capture rich and high-quality vibrational signals from the laryngeal muscles and carotid artery, while high-resolution tokenized segmentation enables users to communicate freely and continuously without expression delays. Additionally, the integration of LLM agents enables intelligent error correction and contextual adaptation, delivering exceptional decoding accuracy (WER < 5%, SER < 3%) and a 55% increase in user satisfaction. The IT thus sets a new benchmark in wearable silent speech systems, offering a naturalistic, user-centered communication aid.

Future efforts in several key areas will guide the continued development of the IT system. First, we are actively expanding our study cohort to include a broader range of dysarthria patients with varying neuromuscular conditions, ensuring that the system is robust across different symptom severities (Table S8). Additionally, we aim to recruit participants from diverse linguistic and ethnic backgrounds to evaluate the system's adaptability across multilingual users. Second, enhancing its linguistic diversity and multilingual support will enable more personalized communication across language barriers, further increasing its accessibility. Third, future iterations will integrate a flexible PCB design, reducing weight and improving conformability to the user's neck. Finally, miniaturizing the system within an edge computing framework will facilitate seamless integration into real-world settings, boosting usability and accessibility.

Looking ahead, the advantages of the IT extend beyond enhancing everyday communication; they contribute to the holistic health of neurological patients, encompassing both physical and psychological well-being. The regained fluency in communication allows patients to re-engage in social interactions, reducing isolation and the associated risk of depression. Moreover, effective communication facilitates real-time, personalized adjustments by rehabilitation therapists, supporting patients' recovery from motor impairments like hemiplegia. Together, these capabilities position the IT as a comprehensive tool for restoring independence and improving quality of life for individuals with neurological conditions.

**Funding:**

National Natural Science Foundation of China 62171014 (SG)

Beihang Ganwei Project JKF-20240590 (SG)

British Council No. 45371261 (LGO)

UK Engineering and Physical Science Research Council (EPSRC) No. EP/K03099X/1 (LGO)

UK Engineering and Physical Science Research Council (EPSRC) No. EP/W024284/1 (LGO)

Haleon through the CAPE partnership contract, University of Cambridge No. G110480 (LGO)

**Author contributions:**

Conceptualization: CT, SG, LGO

Methodology: CT, CL, WY, YJ, XZ

Investigation: CT, CL, WY, YJ, XZ, SL, HM, MX, CW, HY, WW, JC, XF

Visualization: CT, CL, WY, YJ, ZZ

Funding acquisition: SG, LGO

Supervision: CT, SG, LGO

Writing – original draft: CT, CL, WY

Writing – review & editing: CT, SG, SW, XC, NW, PS, YP, WS, MB, LGO

**Competing interests:**

Authors declare that they have no competing interests.

**Data and materials availability:**

The data and code supporting this study will be available from the GitHub repository before publication.

**Supplementary Materials:**

Materials and Methods

Supplementary Text

Figs. S1 to S14

Tables S1 to S8

References

Movies S1 to S2

# Figures

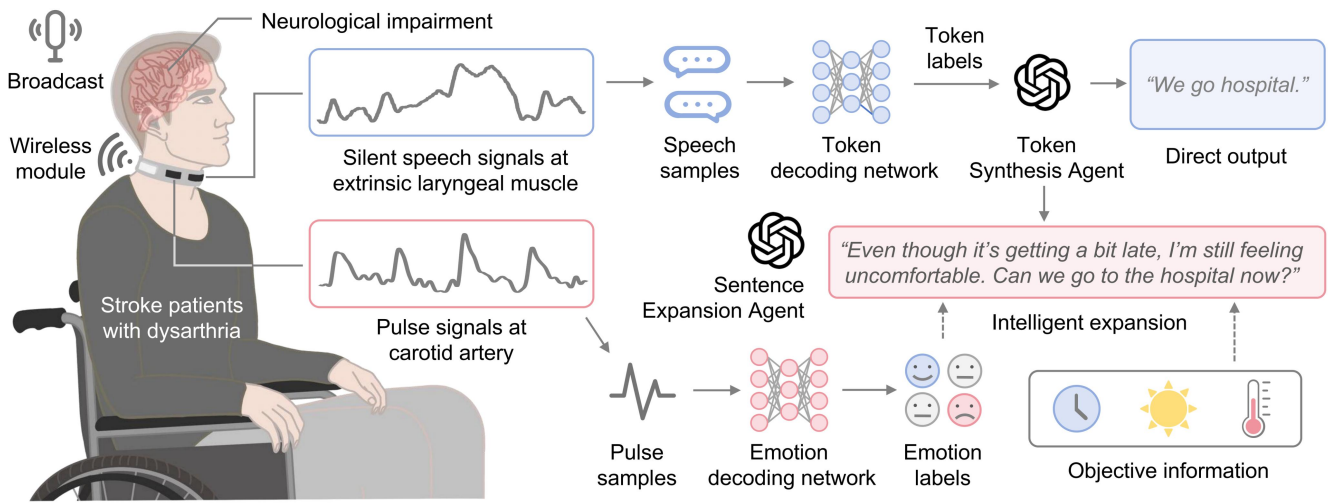

**Figure 1: Schematic of the IT developed for stroke patients with dysarthria.** The system captures extrinsic laryngeal muscle vibrations and carotid pulse signals via textile strain sensors and transmits them to the server through a wireless module. Silent speech signals are processed through a token decoding network, which generates token labels for sentence synthesis. Simultaneously, pulse signals are processed by an emotion decoding network to identify emotional states. The system intelligently integrates both emotional states and contextual objective information (e.g., time, environment) to expand the initial decoded sentences. Through a sentence expansion agent, the decoded output is transformed into personalized, fluent, and emotionally expressive sentences, enabling patients to communicate with a fluency and naturalness comparable to healthy individuals. (Note: Due to grammatical differences between Chinese and English, "We go hospital" is a word-for-word translation of the Chinese expression for "Let's go to the hospital".)

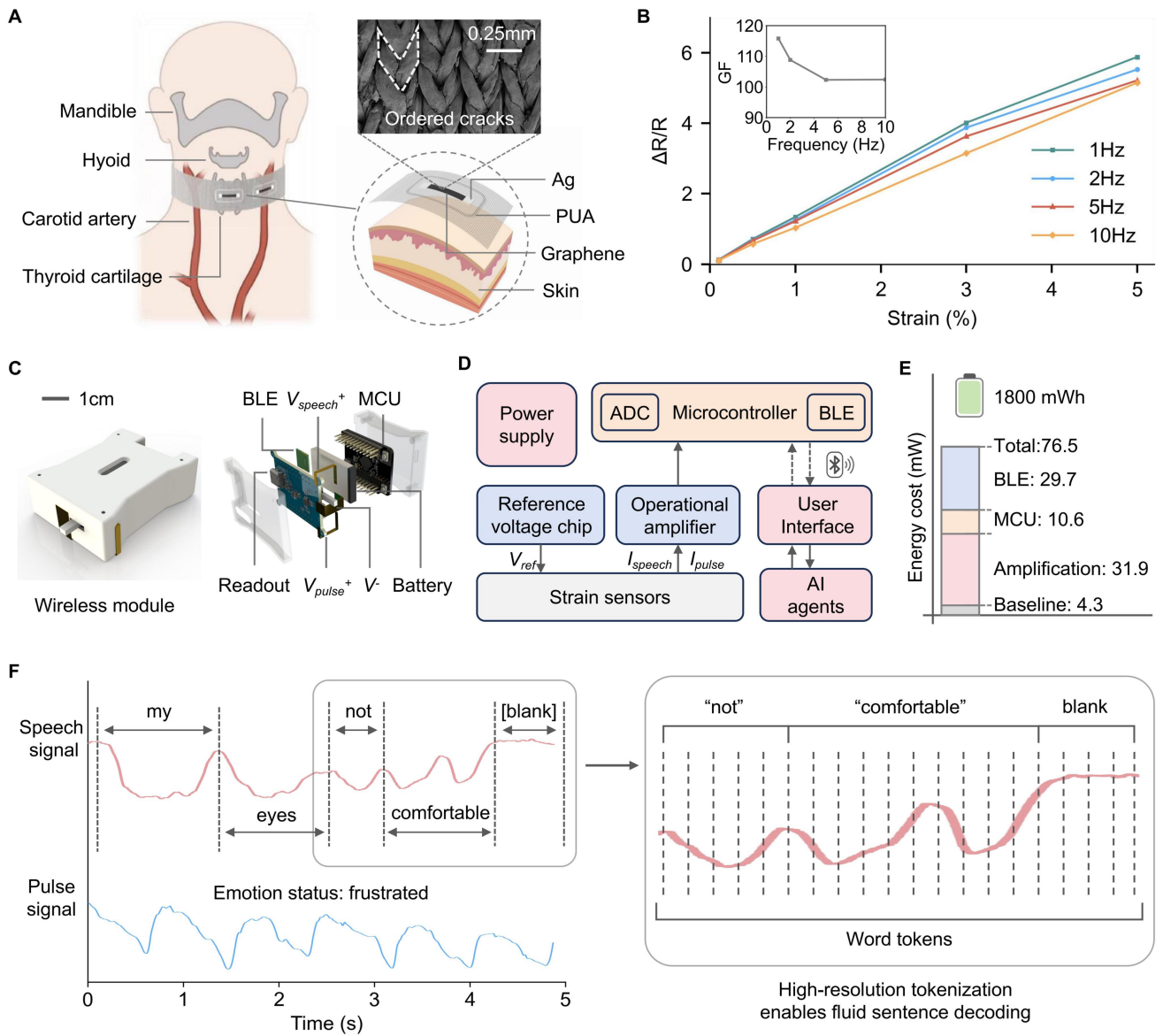

**Figure 2: Hardware and data collection of the IT. A,** Schematic of a textile-based strain-sensing choker. Two channels are aligned with the carotid artery and center of throat, respectively. Each channel consists of a two-terminal crack-based resistive strain sensor surrounded by a polyurethane acrylate (PUA) stress isolation layer. The top right SEM image shows the spontaneous ordered crack structure of the graphene coating. **B,** Relationship between the response to uniaxial stretching (from 0.1% to 5%) and frequency. **C,** Exploded view of the internal components of the PCB. **D,** Diagram of the system communication. **E,** Power consumption of each component during system communication. **F,** Schematic of the high-resolution tokenization strategy.

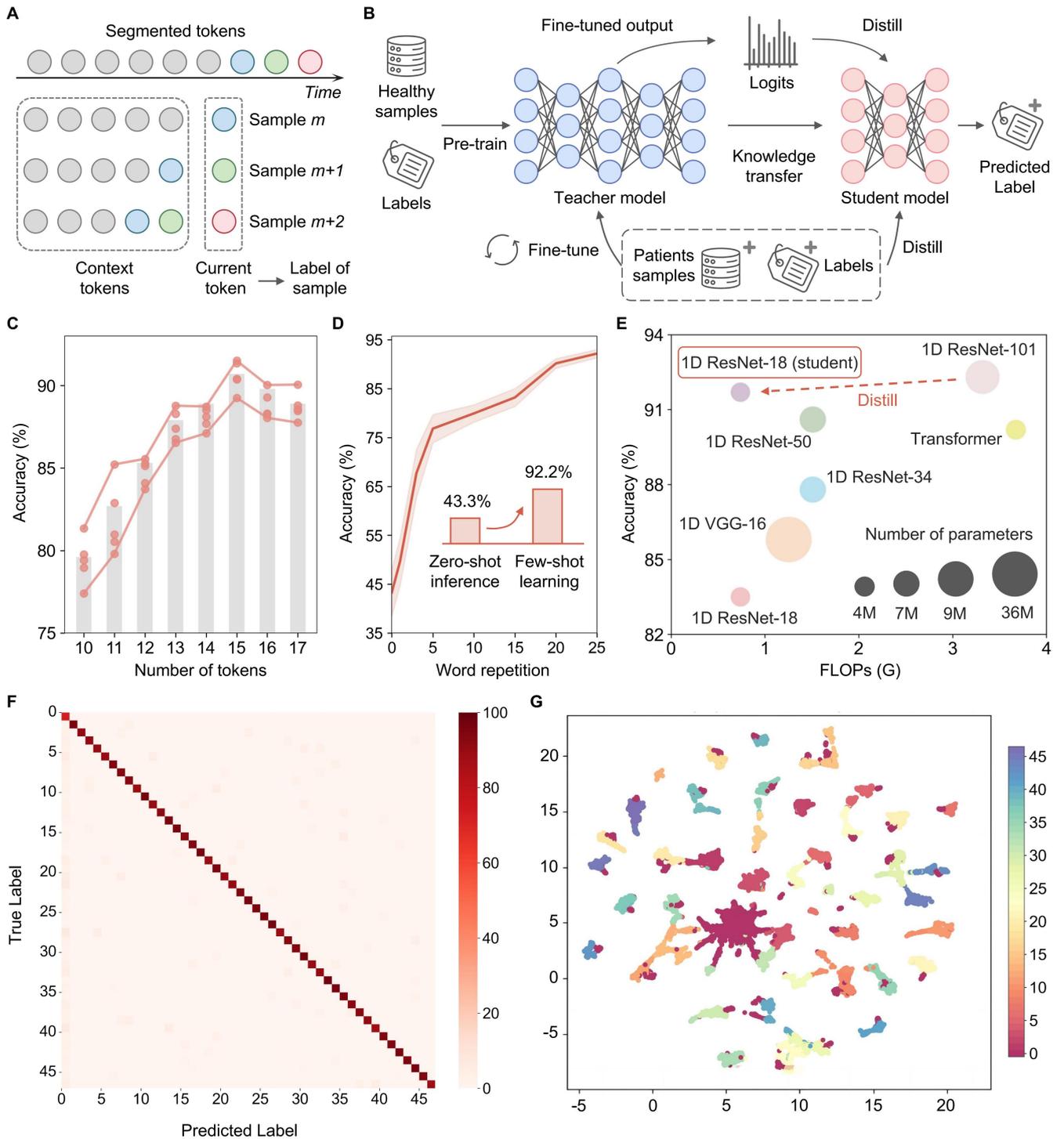

**Figure 3: Token-level decoding framework and performance evaluation. A**, Explicit context augmentation strategy designed to incorporate contextual information by combining tokens into token samples. **B**, Model training pipeline: the teacher model is pre-trained on healthy samples, then fine-tuned on patient samples; knowledge distillation transfers learned features to a student model for efficient prediction. **C**, Comparison of decoding accuracy across different numbers of tokens per sample, showing optimal performance when sufficient contextual information is included. **D**, Accuracy improvement with word repetition in transfer learning process, demonstrating a jump from zero-shot inference (43.3%) to few-shot learning (92.2%) as repetitions increase. **E,** Comparison of model performance across architectures with varying accuracy, FLOPs, and parameter counts; ResNet-101 and ResNet-18 were selected as the teacher and student models, respectively. **F**, Confusion matrix for the final student model. **G,** UMAP visualization of extracted features from the student model, illustrating token clustering patterns that indicate effective decoding and clear separation of different classes.

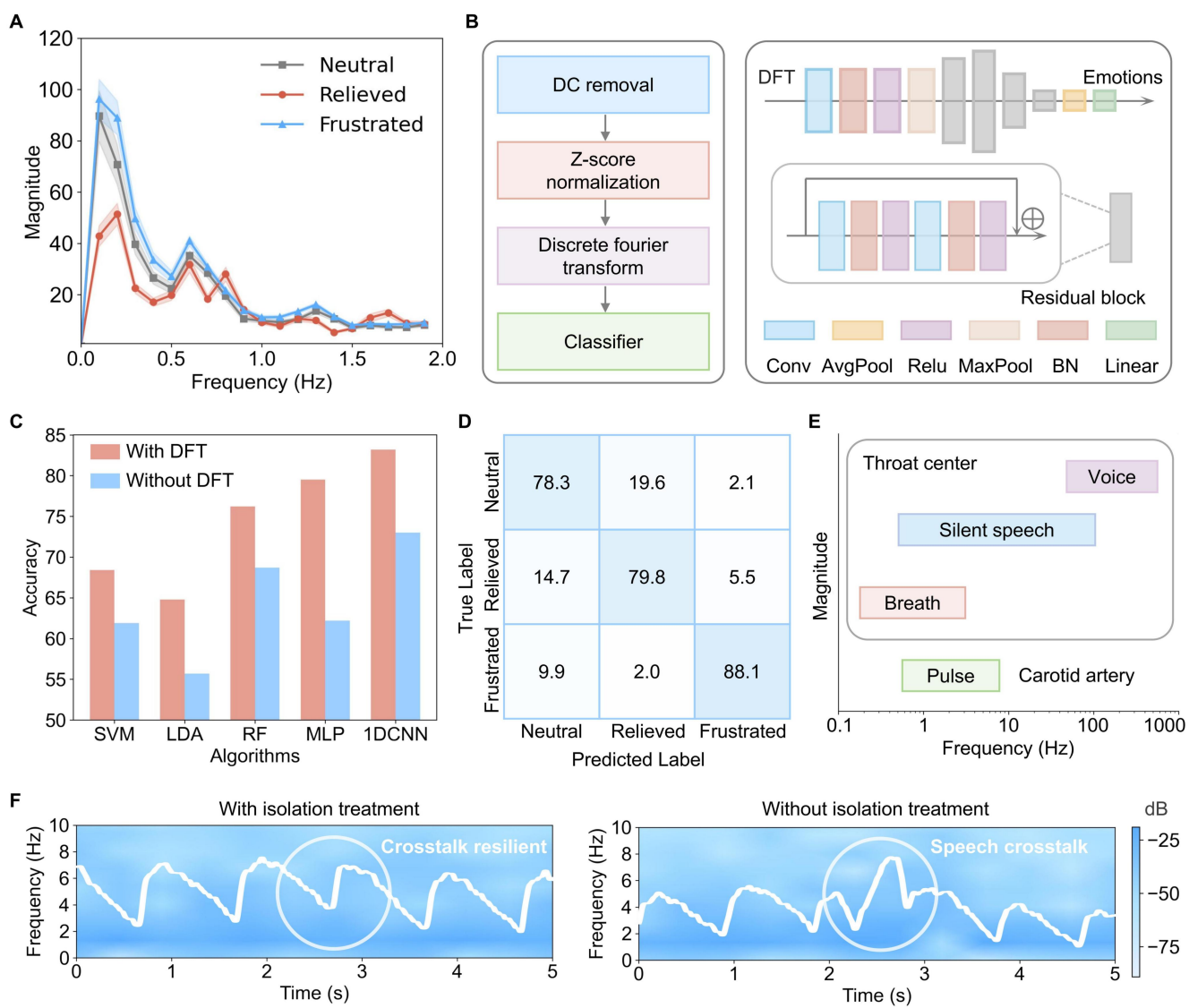

**Figure 4: Emotion decoding framework and performance evaluation. A**, Frequency domain characteristics of carotid pulse signals across three emotional states (Neutral, Relieved, and Frustrated), showing distinct amplitude patterns. **B**, Emotion classification workflow: preprocessing pipeline (left) involving DC removal, Z-score normalization, and discrete Fourier transform (DFT), feeding into a classifier based on a 1DCNN architecture (right) for emotion decoding. **C**, Comparison of classification accuracies across machine learning algorithms (SVM, LDA, RF, MLP, and 1DCNN) with and without DFT preprocessing, highlighting improved performance with DFT. **D**, Confusion matrix for emotion classification. **E**, Frequency and magnitude range of different vibrational signal sources (voice, silent speech, breath, carotid pulse) at neck area. **F,** Time-frequency spectrogram of pulse signals with and without strain isolation treatment when vowel "a" both introduced at 2.5s, demonstrating successful mitigation of speech crosstalk interference after applying the isolation technique.

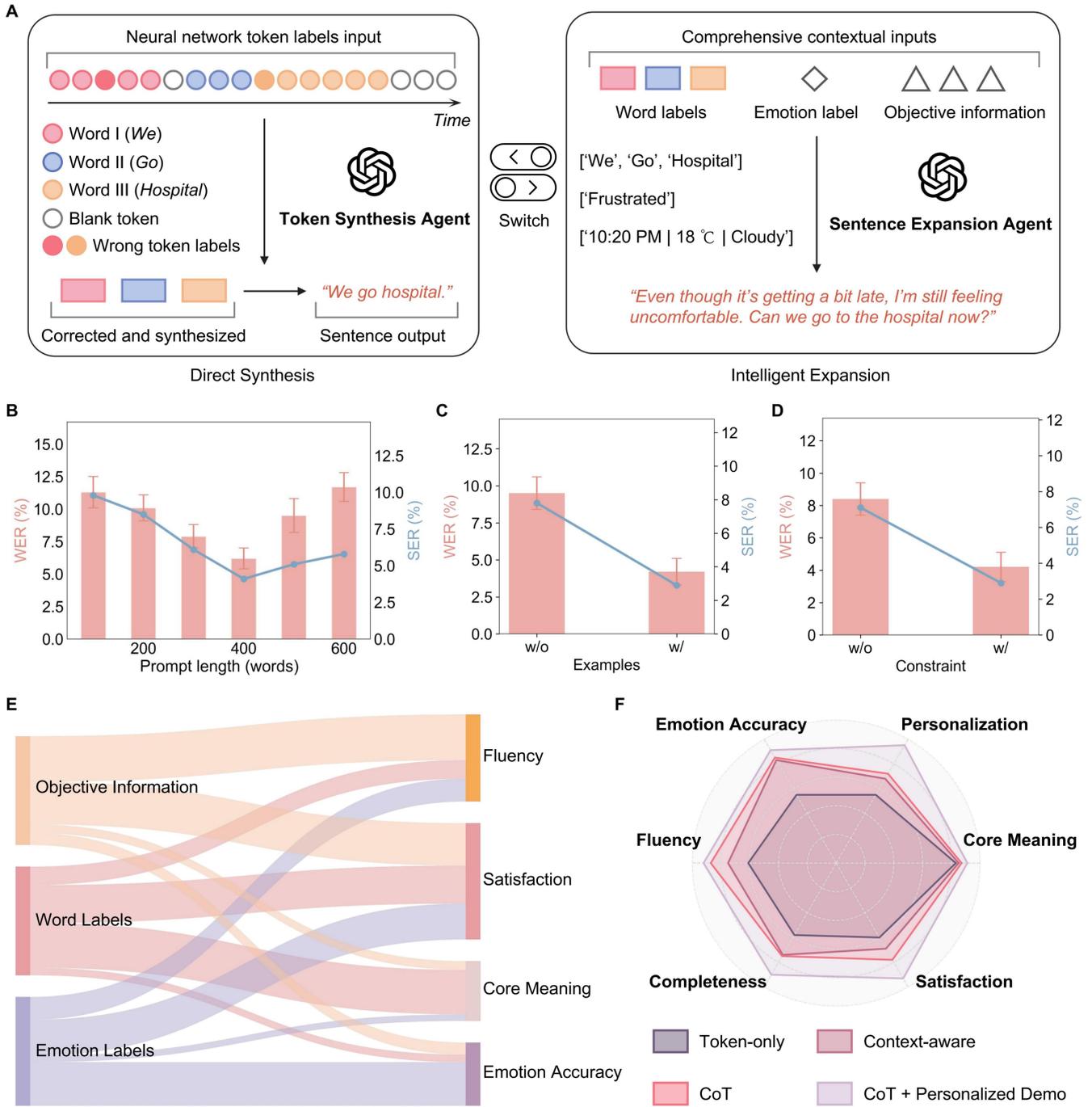

**Figure 5: LLM agents framework and performance evaluation. A**, Schematic of the IT's LLM agents: Token Synthesis Agent (left) directly synthesizes sentences from neural network token labels, while Sentence Expansion Agent (right) enhances outputs with contextual and emotional inputs. **B**, Effect of prompt length on word error rate (WER) and sentence error rate (SER) with optimal performance observed at medium lengths. **C**, Influence of example-based few-shot learning on WER and SER, showing a significant reduction when examples are provided. **D**, Impact of constrained decoding on WER and SER, demonstrating improved accuracy and sentence structure. **E**, Contribution of objective information, word, and emotion labels on key user metrics, including fluency, satisfaction, core meaning, and emotional accuracy (evaluated through ablation experiments). **F**, Radar plot comparing performance across various configurations (Token-only, Context-aware, Chain-of-Thought (CoT), and CoT with personalized demonstration) on fluency, personalization, core meaning, satisfaction, completeness, and emotion accuracy.